\documentclass[prd, aps, reprint, amsfonts, amssymb, amsmath, preprintnumbers, showpacs, nofootinbib, superscriptaddress]{revtex4-1}
\usepackage{graphicx, multirow,soul}
\usepackage[citecolor=blue]{hyperref}
\usepackage[all]{hypcap}
\usepackage{url}
\usepackage{color}
\usepackage{subfigure}
\usepackage{slashed}
\usepackage{float}
\usepackage{bbm}
\usepackage{bbold}

\usepackage{lipsum}

\newcommand{\figref}[1]{Fig.~\ref{#1}}

\usepackage{tikz}
\tikzset{node distance=2cm, auto}

\newcommand{\bq}{\begin{eqnarray}}
\newcommand{\nq}{\end{eqnarray}}

\begin{document}

\title{Teaching Quantum Computing to High School Students}

\author{Ciaran Hughes}
\affiliation{Theoretical Physics Department, Fermi National Accelerator Laboratory, Batavia, IL 60510, USA.}
\author{Joshua Isaacson}
\affiliation{Theoretical Physics Department, Fermi National Accelerator Laboratory, Batavia, IL 60510, USA.}
\author{Anastasia Perry}
\affiliation{Illinois Mathematics and Science Academy, Aurora, IL 60506, USA.}
\author{Ranbel Sun}
\affiliation{Phillips Academy Andover, Andover, MA 01810, USA.}
\author{Jessica Turner}
\affiliation{Theoretical Physics Department, Fermi National Accelerator Laboratory, Batavia, IL 60510, USA.}

\date{\today}

\begin{abstract}
Quantum computing is a growing field at the intersection of physics and computer science. The goal of this article is to
    highlight a successfully trialed quantum computing course for high school students between the ages of 15 and 18
    years old. This course bridges the gap between popular science articles and advanced undergraduate textbooks.
    Conceptual ideas in the text are reinforced with active learning techniques, such as interactive problem sets and
    simulation-based labs at various levels. The course is freely available for use and download~\cite{Perry:2019bqg} under the Creative Commons ``Attribution-
NonCommercial-ShareAlike 4.0 International'' license. 
\end{abstract}
\preprint{FERMILAB-PUB-20-151-T} 
\maketitle

Popular media often portrays quantum computing as a life changing technology~\cite{8585034}. For example, quantum
computing could have the power to make our bank accounts vulnerable to hacking, design life-saving drugs, and prevent
cyber attacks from Nation-State actors. For these reasons, many countries are investing billions of dollars to successfully develop this disruptive technology.

\begin{figure}[h!]\label{fig:overview}
\includegraphics[width=0.45\textwidth, keepaspectratio,]{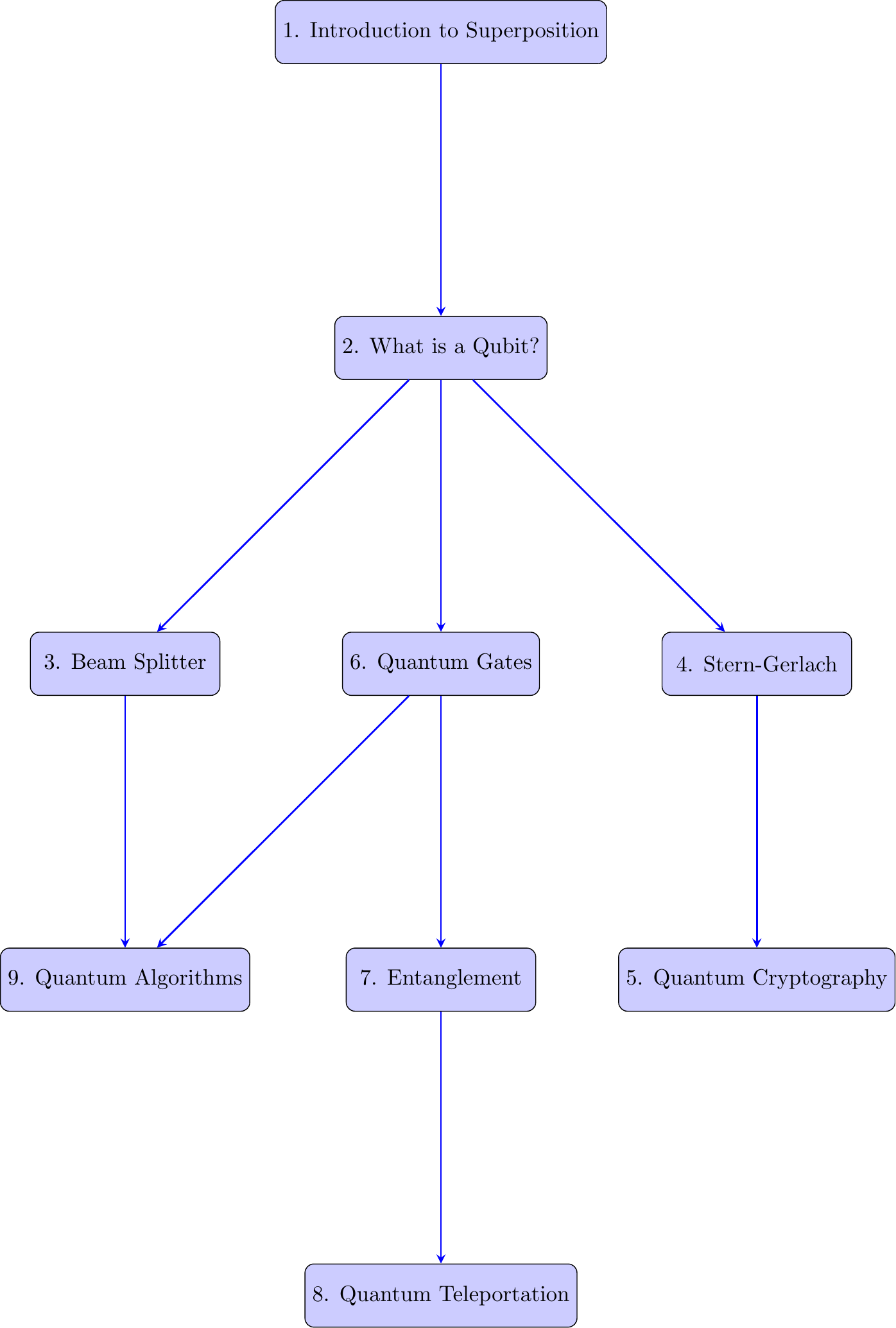}
\caption{Flowchart of learning outcomes}
\end{figure}
\begin{figure*}[t!]\label{fig:data}
\includegraphics[width=\textwidth]{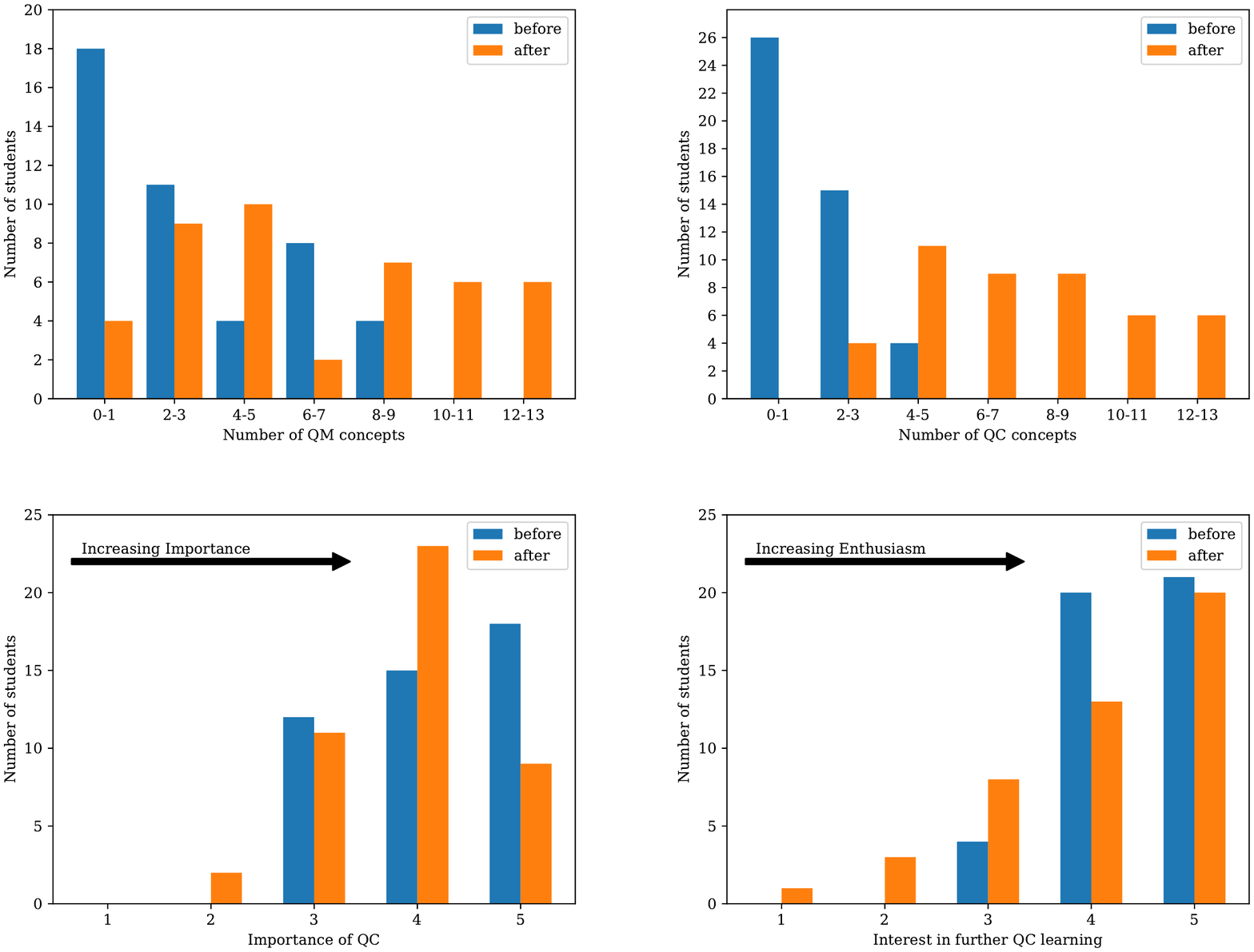}
\caption{The top left  (right)  plot shows the number of quantum mechanic (computing) concepts the students knew before (blue) and after (orange) taking the module. The bottom left plots shows the perceived importance of quantum computing before (blue) and after (orange) taking the module. The bottom right plot shows the distribution of interest in learning more quantum computing (5 indicates the maximum level of interest).}
\end{figure*}

The  necessary foundations for learning quantum computing is quantum physics. H.~K.~E.~Stadermann, E.~van den Berg, and M.~J.~Goed~\cite{ED} analyzed high school quantum physics curricula (a necessary component for quantum computing) from 15 countries. A list of important topics related to quantum physics that are covered at high school level was compiled, and the importance of the nature of science highlighted.  Most notably, the USA was not included in the analysis because quantum physics is currently not taught widely in high schools. As such, the future competitiveness of the USA in the realm of quantum technology is reduced. This quantum computing course aims to help develop the future workforce and disseminate knowledge at an accessible level in regards to quantum computing.

More concretely, the course ``Quantum Computing as a High School Module'' has been  designed for high school students between 15 and 18 years old, but is also useful for community colleges or undergraduates studying subjects outside of physics, such as computer science, engineering, or mathematics. Two of the authors are experienced high-school teachers, which ensures that the material is at the appropriate level for the intended demographics. Three of the authors are scientists at a U.S. national laboratory, which ensures the material is at the cutting-edge and forefront of innovation. 
		
The material assumes basic knowledge of electricity, magnetism, and waves from high school-level physics. Introductory
modern physics (photoelectric effect, wave-particle duality) is helpful but not required. Computer programming
experience is not required, as the course covers this aspect where needed, for example on how to use the IBM quantum
computers\footnote{See Section 10.4 of the course~\cite{Perry:2019bqg} for details.}. The material starts by explaining
a few of the basic concepts which are required to understand quantum computing: superposition of particles, quantum
measurement, and entanglement.  The course was designed with three different levels of complexity depending on the
students' experience with mathematics and abstract reasoning. All sections and problems are labeled according to these
difficulty levels\footnote{See Section 0.2 of the course~\cite{Perry:2019bqg} for details on the different levels.}. In addition, the intermediate and advanced sections within each chapter are labeled such that students and teachers can skip them if necessary. The topics that are covered are shown in \figref{fig:overview}, and the units are best studied in the order shown. For those with limited time, \figref{fig:overview} outlines the minimum recommended prerequisites for each unit. Furthermore, it is possible to skip different chapters while leaving the core content fully understandable.

The module finishes with  particularly interesting topics such as quantum algorithms, quantum cryptography, and quantum teleportation. The last chapter contains interactive worksheets that the students particularly enjoyed. These include videos, simulations, and a walkthrough of Schrodinger's worm (analogous to the cat) which is run on the IBM quantum computer~\cite{IBM}. In addition, the BB84 quantum encryption group exercise shows how a quantum computer can protect against hacking. Such activities are designed to help students digest the information using an active-learning paradigm. From the classroom experience, these activities were crucial for student engagement and knowledge retention.
\begin{figure}[h!]\label{fig:cloud}
\includegraphics[width=0.5\textwidth]{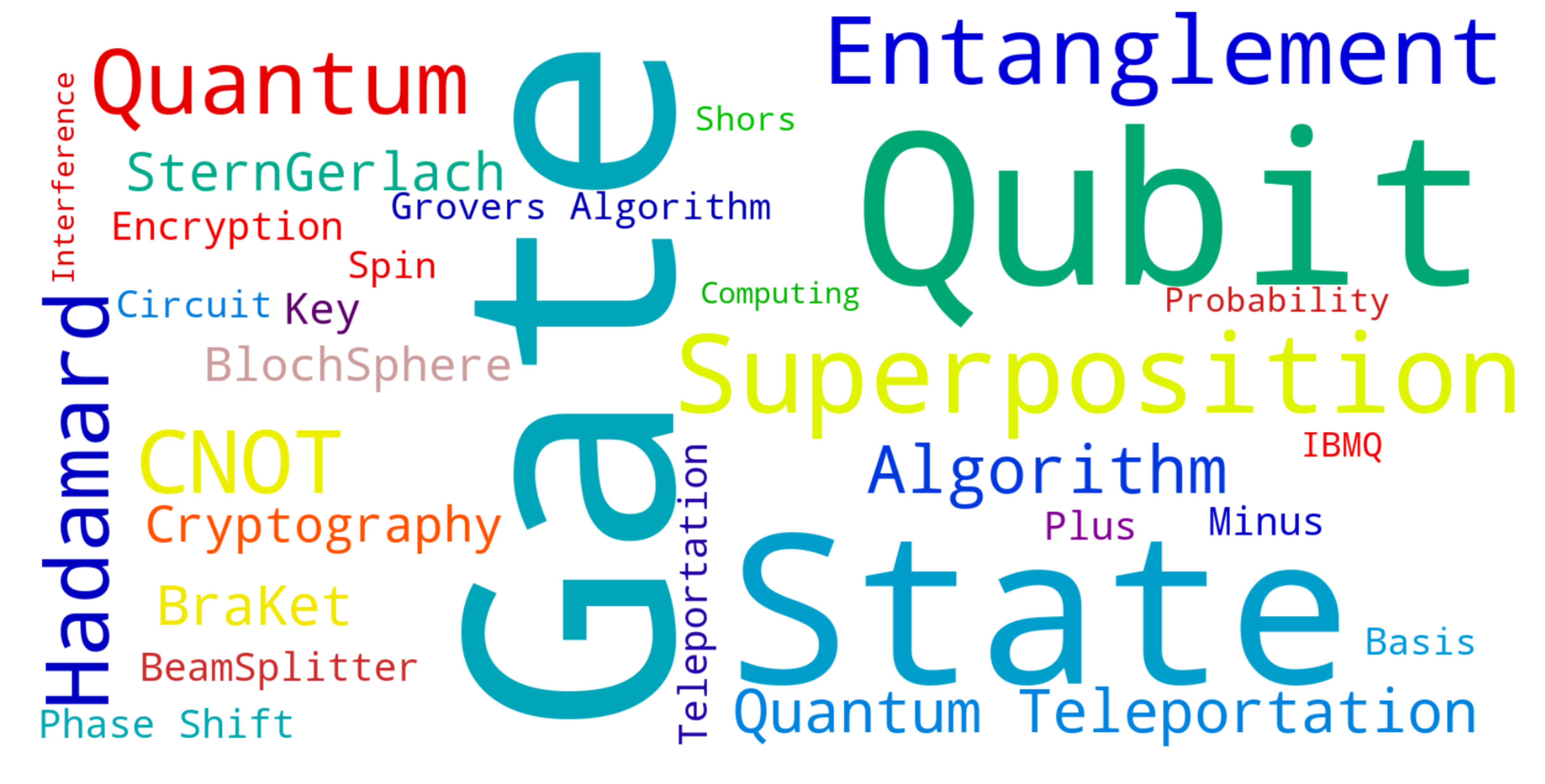}
\caption{Word cloud of QC concepts learned after finishing the course. }
\end{figure}
Note that the module is not designed to be a comprehensive introduction to modern physics. Rather, it focuses on topics that students may have heard about through popular media but that are not typically covered in a general high school course. Given the usual constraints on teaching time, these materials could be used after the advanced physics (AP) examinations, in an extracurricular club, or as an independent project resource to give students a taste of quantum computing. 
In the following, two ways that the course was taught are discussed, and data is presented demonstrating the course's effectiveness. 
	
One of the authors taught a class consisting of 25 students (17 sophomores, 7 juniors, 1 senior). Additionally, two
senior students who had prior QC knowledge assisted with the class. Some of the students had taken physics classes while
others had not. The module was completed in five days, consisting of four 3.5 hour sessions and an additional one hour
session. Due to the diversity of the class, a traditional mode of teaching was adopted: the first 30--40 minutes consisted of a lecture, followed by group activities, and then a class discussion. The students used the module for reference and as a source for the activities. 

For the activities, the students were split into groups with mixed and complementary abilities. The mix was determined from the use of surveys, which were designed to help determine the students familiarity with quantum mechanics and computing~\cite{surveys}. By assessing students' strengths, it was possible to present the material in a manner appropriate to the students abilities and interests. For example, many of the students in the class had a strong mathematics and computer science background but rudimentary physics knowledge. These students were able to engage with the mathematical and programming activities well but struggled with the concepts related to quantum mechanics such as the Stern-Gerlach apparatus, beam splitters, and phase shifts. Therefore, more focus was given to the mathematical formulation which aided the understanding of quantum computing. The knowledge of students' strengths should be used to guide teachers in their choice of topics, which are outlined in  \figref{fig:overview}.
\begin{figure}[t!]\label{fig:students1}
\includegraphics[width=0.45\textwidth,keepaspectratio]{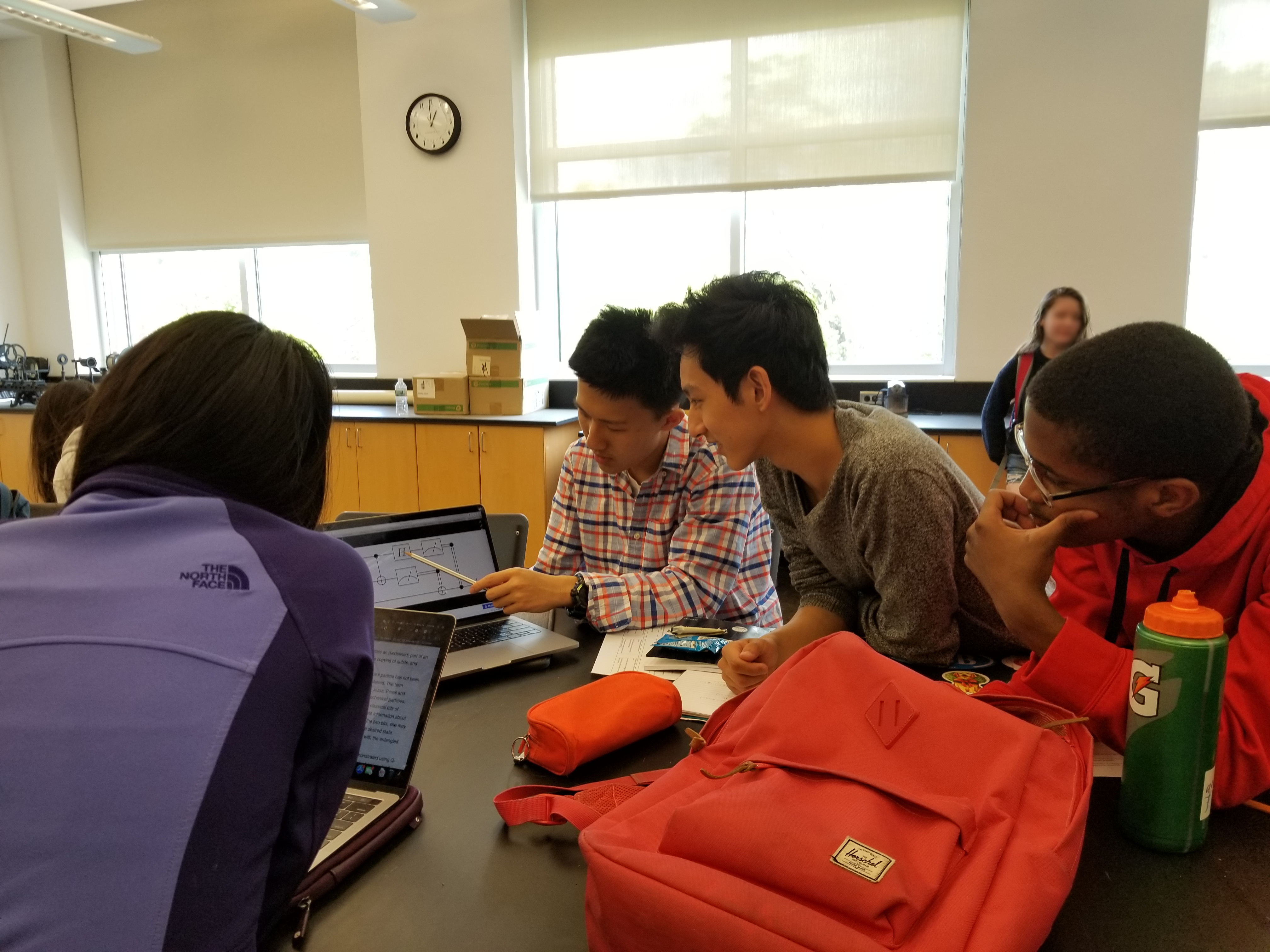}
\caption{Students building the circuit for quantum teleportation.}
\end{figure}
\begin{figure}[t!]\label{fig:students2}
\includegraphics[width=0.4\textwidth,keepaspectratio]{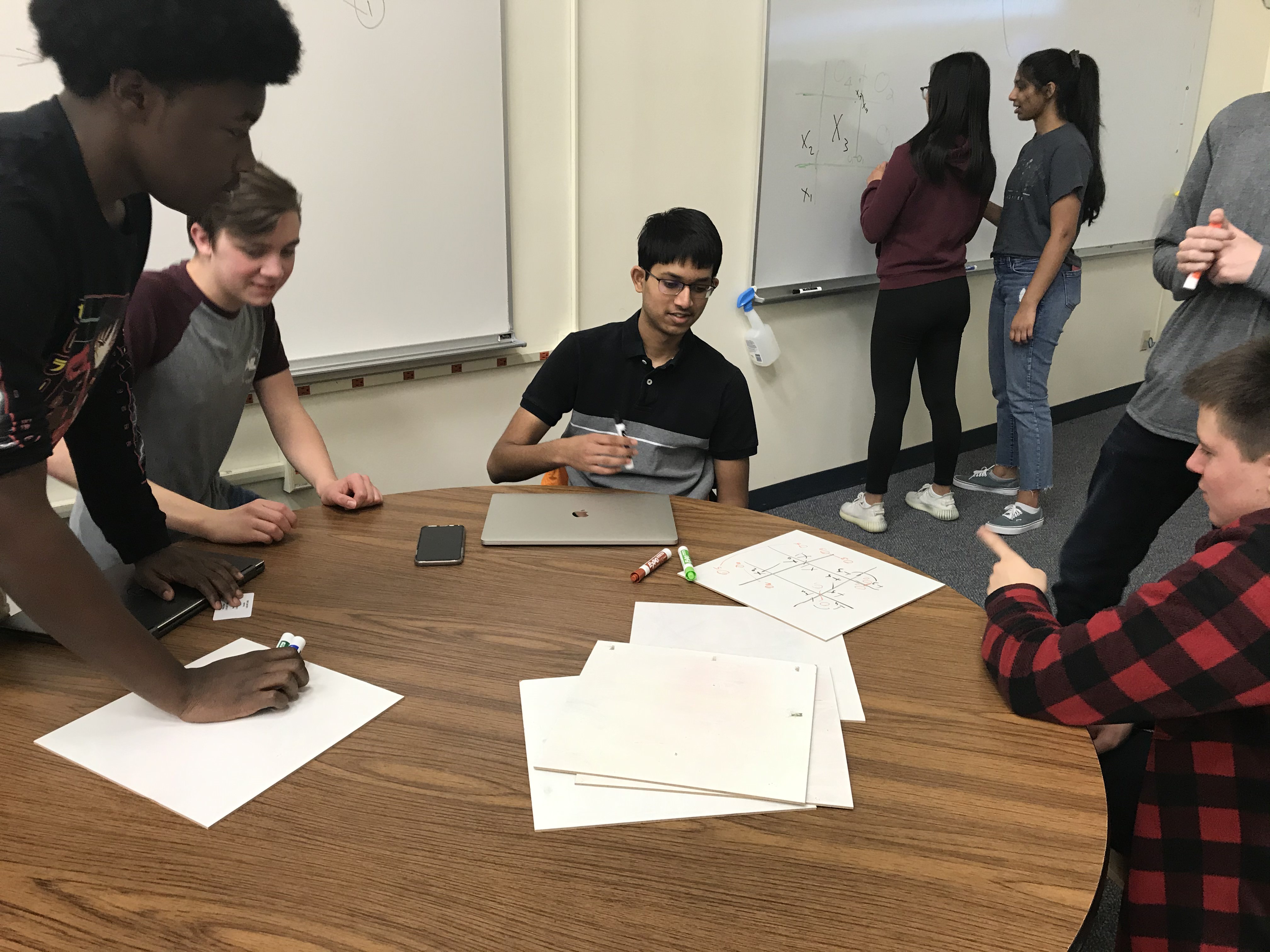}
\caption{Students learning about quantum entanglement through quantum tic-tac-toe.}
\end{figure}

Another one of the authors taught the course to 20 senior students in a ``flipped classroom'' active-learning paradigm~\cite{active}: the students read the module outside of class and performed activities during the teaching sessions. This module was taught after an introductory calculus-based physics course (AP Physics C). The students were split into smaller groups and each choose their own topic from the course (the different topics are shown in \figref{fig:overview}). One motivation for this self-taught setup was to determine if the module was appropriate as a standalone resource. This proved to be true, as the students demonstrated their understanding when engaging in the activities and answering the questions. Given the interactive nature of the sessions, the students were particularly engaged, delegated questions to each other, asked for clarification from each other, and brainstormed conceptual understanding.

The students had positive experiences with the interactive aspects of the course, in particular on how to use a real quantum computer through IBM's website~\cite{IBM}. In addition, the interactive quantum encryption activity, quantum tic-tac-toe, and example problems were highlighted multiple times as positive memorable experiences.  \figref{fig:students1} and \figref{fig:students2} show students interacting with these activities in the classroom. The 45 students were asked to fill out a survey immediately before and after finishing this course. As part of the surveys, the students were asked to list as many quantum mechanics (QM) or computing (QC) concepts that came to mind. This data allowed the course to be tailored to their base knowledge and also provided a simple way to measure how much the students learned from the course.  
	 
The top left and right histograms of \figref{fig:data} show that after taking the course, students significantly improved their QM and QC knowledge base, indicated by the orange distributions shifting to the right. By taking the QC concepts used to make the top-right histogram,  \figref{fig:cloud} shows this data as a word cloud illustrating the most common QC concepts which the students learned. Note that \figref{fig:cloud} does not show any previously known concepts prior to taking the course, as there were too little to be illustrative.  Furthermore, the enjoyment of the students and the effectiveness of the module in knowledge retention is evident from the student feedback and assessments. 

The distributions of perceived societal importance of QC are shown in the bottom left plot of \figref{fig:data}. Using
our metric where 1 (5) indicated the minimal (maximal) importance of QC, the median of the before and after data is 4.
The change in the distributions highlights that the students appreciated that QC technology is in its infancy, and much
of the ``hype'' generated by popular media is hyperbole. Feedback from the students supports this understanding:
``\emph{Sci-fi and science magazines make quantum computing seem more useful than it currently is}\ldots'' This is not to undermine the enormous potential of QC, and many of the students showed appreciation for the ongoing research efforts: ``\emph{I had zero knowledge about quantum computing before this project, and I now have a broad understanding of why so many scientists are invested in quantum computing.}''

The general enthusiasm of students in pursuing further QC learning is shown in the bottom right plot of \figref{fig:data}. Prior to completing course, the median was 4, and this did not change after completion. The change in distribution demonstrates that students successfully learned about QC and some wished to continue studying QC further at university while others learned enough to satisfy their curiosity. This also highlights the effectiveness of the module.

The main obstacle for new course utilization, for both teachers and students, is time. From the perspective of the teacher, upskilling to learn the course content may be required. This objective can be achieved in numerous ways. One highly recommended method is attending a fully compensated teacher training workshop through the AAPT~\cite{AAPT}, and to initially focus on one stream of the course rather than the entire module. In addition, one particularly appealing aspect of quantum computing is its interdisciplinary nature which requires mathematics, physics, and computer science. This would be an excellent module for collaborative teaching. One of the teachers was aided by two senior students who had undertaken summer internships, and these students proved to be useful in explaining key concepts and helping with class activities. For enthusiastic students eager to learn more 
there are excellent facilities: programming on IBM's quantum devices is available on their qiskit website~\cite{qiskit} along with instructional YouTube videos~\cite{YT} and  IBM internships are available for  students~\cite{intern}.

From the students perspective, it is important to specify a course stream that takes an appropriate amount of time. Based on \figref{fig:data}, a one-week module is found to be empirically  useful and impactful. Nonetheless, students would have preferred a longer study time, and one student suggested to ``\emph{replace special relativity with QC next year.}'' Another possibility would be to provide this module as a summer course for interested students.

``Quantum Computing as a High School Module'' has been demonstrated to be successful at bridging the gap between popular media and undergraduate texts, generating enthusiasm for studying STEM, and ensuring the nations future workforce remains competitive. It is the hope of the authors that teachers find this resource useful to teach quantum physics/computing, and we encourage teachers or students to contact us regarding the course.

The following article has been submitted to ``The Physics Teacher''. After it is published, it will be found \href{https://publishing.aip.org/resources/librarians/products/journals/}{here}.

\bibliographystyle{apsrev4-1}
\bibliography{QC}{}
\end{document}